\documentclass[showpacs,pra,aps]{revtex4}
\usepackage{dcolumn}
\usepackage{bm}
\usepackage{graphicx}

\begin{document}

\title{The Theory of Two-Photon Interference in an EIT System}
\author{Jianming Wen and Morton H. Rubin}
\email{rubin@umbc.edu}
\affiliation{Physics Department, University
of Maryland, Baltimore County, Baltimore, MD 21250}

\date{\today}

\begin{abstract}
We examine the possibility of storing and retrieving a single
photon using electromagnetically induced transparency (EIT). We
consider the theory of a proof of principle two-photon
interference experiment, in which an atomic vapor cell is placed
in one arm of a two-photon interferometer. Since the two-photon
state is entangled, we can examine the degree to which
entanglement survives. We show that while the experiment might be
difficult, it should be possible to perform. We also show that the
two-photon interference pattern has oscillatory behavior.
\end{abstract}


\maketitle

\section{Introduction}

It is well-known that processes linear in the electromagnetic field have
propagation properties that are independent of the strength of the field.
This is part of the definition of linearity. In the quantum mechanical case,
this is understood because the coupling between the electromagnetic field
and matter is determined by the modes of the electromagnetic field
independent of the state of the field, that is, independent of how the modes
are populated. Consequently, the question of whether an experiment can be
realized at the single photon level often reduces to a detailed analysis of
the experiment. An example of this is the question of whether it is possible
to coherently store and retrieve light at the single photon level. In this
paper, we analyze this question by studying two-photon interference in which
an atomic vapor cell is placed in one arm of the interferometer and the cell
operated under the conditions of electromagnetically induced
transparency (EIT) \cite{Harris,Mara,ScuZus}.

The essence of EIT is to create destructive interference of the transitions
for a three-level system in order to control the optical responses of the
system. Harris \textsl{et al.} \cite{HFK} first suggested how EIT can be
used to slow the speed of light significantly compared with the vacuum case.
Early experiments \cite{HauH,SL,Kash,BKRY} on slow light have demonstrated
that the group velocity $v_{g}$ can be reduced to several meters per second.
The results reported in these experiments are based on the fact that EIT not
only makes absorption zero at the resonant situation but also leads to a
rapidly changing dispersion profile. The condition for slow light
propagation leads to photon switching at an energy cost of one photon per
event \cite{HarYama} and to efficient nonlinear processes at energies of a
few photons per atomic cross section \cite{HarHau}.

Theoretically there are two ways to implement EIT. One way is adiabatic EIT
\cite{Harris} in which both the probe and coupling resonant lasers are
adiabatically applied. After the system evolves into a steady state, EIT
occurs for arbitrary intensities of the probe and coupling lasers. The other
way is the transient-state EIT \cite{ScuZus}, where resonant probe and
coupling lasers are simultaneously applied. EIT occurs only when the
intensity of the probe laser is much weaker than that of the coupling one.

In EIT the system is driven by two fields called the coupling and the probe,
see Fig. 1 for notation. Most of the early theory and experiments of EIT
took both coupling and probe lasers as classical external fields, see
however \cite{Field}. A disadvantage of the this approach is that it is hard
to deal with atom-photon and photon-photon quantum entanglement which is of
importance not only because of the interest of fundamental physics, but also
for their potential applications to quantum computation and quantum
communication \cite{NeilChuang}. Recently a number of papers have treated
the probe laser quantum mechanically \cite{FYL,FLukin}.

In this paper we will take as the probe source a photon produced
using spontaneous parametric down conversion (SPDC)
\cite{Klyshko,Shih}. In this nonlinear optical process a
high-frequency photon is annihilated and two lower frequency
photons, conventionally referred to as the signal and idler, are
generated. The pair of photons are entangled in frequency and wave
number. The correlations of the entangled two-photon system can be
measured by means of coincidence counting detection. The purpose
of the present paper is to study the optical properties of a
transient-state EIT system interacting with one quantized field
and investigate the response of the EIT medium to the nonclassical
light field. We determine the two-photon interference in which one
of the photons is stored and released. We want to examine whether
the two-photon entangled state can be preserved in the process.
There is an inherent mismatch of four orders of magnitude between
the spectral bandwidth of SPDC and very narrow bandwidth of
EIT; however, we shall see that it still may be possible to
perform a proof of principle experiment.

This paper is organized as follows. In Sec. II, we discuss the quantum
mechanical description of the system Hamiltonian. The equations of
motion for the quantum probe field is given. In Sec. III, we investigate the optical
properties of the EIT system related to the proposed experiment. In Sec. IV, the two-photon interference
experiment of the correlated photon pairs generated by SPDC will be
discussed when the signal photon is delayed in an EIT system. Finally, we
summarize our results in Sec. V.

\section{Evolution of the operator}

To describe the interaction of electromagnetic fields, the
standard method is to start with the Bloch equations for the
atomic density-matrix elements under the adiabatic assumption and
moderate intensities of the fields. The equation of motion of the
fields, which are treated as classical fields, are obtained from
Maxwell equations. The detailed discussion can be found in
\cite{FLukin,MLukin}.

\begin{figure}[tbp]
\includegraphics{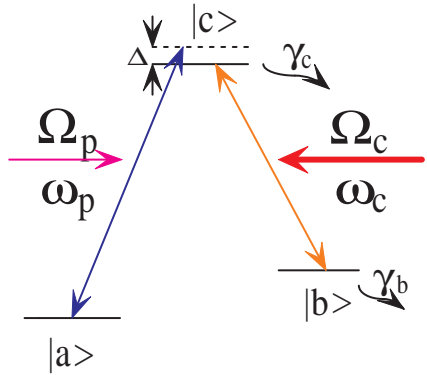}
\caption{$\Lambda $-type atomic configuration of EIT. The strong
coupling field with frequency $\omega _{c}$ (Rabi frequency
$\Omega _{c}$) resonantly drives the states $|b\rangle\rightarrow
|c\rangle$ and the weak probe field with frequency $\omega _{p}$
(Rabi frequency $\Omega _{p}$) is applied to the states
$|a\rangle\rightarrow |c\rangle $ with a small frequency detuning
$\Delta $, respectively. $\gamma _{c}$ and $\gamma _{b}$ are decay
rates of states $|c\rangle $ and $|b\rangle $.} \label{Fig.1}
\end{figure}

We consider a three-level atom with $\Lambda $-type configuration
interacting with two electromagnetic fields, which is shown in Fig. \ref
{Fig.1}. The two lower metastable levels $|a\rangle $ and $|b\rangle $ are
coupled to the upper excited level $|c\rangle $
by the probe and coupling fields, with Rabi frequencies $\Omega _{p}$ and $%
\Omega _{c}$, respectively. The coupling field is taken to be resonant $%
\omega _{cb}=\omega _{c},$ and the probe field is detuned by $\Delta =\omega
_{ca}-\omega _{p}$(see Fig. \ref{Fig.1}). In the interaction picture, the
Hamiltonian of the system can be expressed as
\begin{equation}
\hat{H}=\hbar \left(
\begin{array}{clrr}
0 & \Omega _{p}^{*} & 0 &  \\
\Omega _{p} & \Delta -i\gamma _{c} & \Omega _{c}^{*} &  \\
0 & \Omega _{c} & \Delta -i\gamma _{b} &
\end{array}
\right)  \label{Hamiltonian}
\end{equation}
in the basis $\{|a\rangle ,$ $|c\rangle ,$ $|b\rangle \}.$ Assuming $|\Omega
_{p}|<<|\Omega _{c}|$, in the steady-state approximation, the eigenvalue of $%
\hat{H}$ associated with EIT is, to leading order in $|\Omega
_{p}|^{2}$,
\begin{equation}
\zeta = -\hbar \frac{|\Omega _{p}|^{2}(\Delta -i\gamma _{b})}{(\Delta
-i\gamma _{c})(\Delta -i\gamma _{b})-|\Omega _{c}|^{2}}  \label{Eigenvalue}
\end{equation}
where $\zeta $ reduces to zero when $|\Omega _{p}|\rightarrow 0$. This is
the energy of the probe transition from the initially prepared atomic state $|a\rangle $.
The polarization of
a medium with atomic density $N$ is the partial derivative of the
interaction Hamiltonian with respect to the amplitude of the electric field,
 i.e.,
\begin{equation}
P=-N\langle \frac{\partial {H}}{\partial {E^{*}}}\rangle \ =-\frac{N\mu ^{*}%
}{\hbar }\langle \frac{\partial {H}}{\partial {\Omega ^{*}}}\rangle
\label{Polarization}
\end{equation}
where the second term comes from the definition of the Rabi frequency and $%
\mu $ is the electric dipole matrix element of corresponding transition.

In the weak-probe limit, the excited states will have a very small
population, and the system evolves adiabatically so most of atoms are in the
initially prepared state $|a\rangle $. Under this condition, the interaction
Hamiltonian $\hat{H}$ shown in Eq. (\ref{Polarization}) can be replaced by
its eigenvalue, therefore,
\begin{equation}
P=-\frac{N\mu ^{*}}{\hbar }\langle \frac{\partial {\zeta }}{\partial {\Omega ^{*}}}%
\rangle.  \label{Polarization1}
\end{equation}
From Eq. (\ref{Eigenvalue}), one can obtain the polarization of the EIT
medium at the probe frequency $\omega _{p}$,
\begin{equation}
P_{p}(\omega _{p})=\frac{N|\mu |^{2}}{\hbar }\frac{(\Delta -i\gamma _{b})}{%
(\Delta -i\gamma _{c})(\Delta -i\gamma _{b})-|\Omega _{c}|^{2}}E_{p},
\label{ProbeP}
\end{equation}
and from $P_{p}=\epsilon _{0}\chi E_{p}$, one finds the linear
susceptibility,
\begin{equation}
\chi (\omega )=\frac{N|\mu |^{2}}{\hbar \epsilon
_{0}}\frac{(\Delta -i\gamma _{b})}{(\Delta -i\gamma _{c})(\Delta
-i\gamma _{b})-|\Omega _{c}|^{2}}. \label{suscept}
\end{equation}
This result can be found in \cite{ScuZus}.

To quantize the probe field, we shall assume that the field is a
quasi-monochromatic wave traveling in the $z$-direction
\begin{equation}
\hat{E}_{p}=\sqrt{\frac{\hbar \omega _{p}}{2\epsilon _{0}V_{Q}}}\left( \hat{a%
}_{p}(t)e^{i(k_{p}z-\omega _{p}t)}+\hat{a}_{p}^{\dagger
}(t)e^{-i(k_{p}z-\omega _{p}t)} \right). \label{quantizedProbe}
\end{equation}
One can introduce the effective Hamiltonian operator which describe the
interaction between the fields and the EIT medium:
\begin{mathletters}
\begin{equation}
\hat{H}_{eff}=\frac{\hbar \omega _{p}}{2}\chi \hat{a}_{p}^{\dagger }(t)\hat{a%
}_{p}(t).  \label{interactionH}
\end{equation}
The evolution of the annihilation operator $\hat{a}_{p}$ is given by
\end{mathletters}
\begin{equation}
\frac{\partial \hat{a}_{p}}{\partial t}=-i\frac{\omega _{p}}{2}\chi [\hat{a}%
_{p}^{+}\hat{a}_{p},\hat{a}_{p}]+\hat{F}=i\frac{\omega _{p}}{2}\chi \hat{a}%
_{p}+\hat{F}  \label{eq motion a}
\end{equation}
where we must include the noise operator $\hat{F}$ \cite{ScuZus}. Since it
will not contribute to the counting rates, we will drop it. If the
interaction starts at $t=0$, field at the output of the EIT cell is given by
\begin{equation}
\hat{a}_{p}^{(out)}(t)=e^{-\frac{\chi ^{\prime \prime }\omega _{p}t}{2}}e^{i%
\frac{\chi ^{\prime }\omega _{p}t}{2}}\hat{a}_{p}^{(in)}(0)  \label{output}
\end{equation}
where we have written $\chi =\chi ^{\prime }+i\chi ^{\prime \prime }$ 
in terms of its real and imaginary parts $\chi ^{\prime }$ and $\chi ^{\prime \prime }$, respectively. The
imaginary part of the susceptibility describes the amplitude
change and the real part of the susceptibility gives the phase
shift of the operator. Using $t=z/c$ Eq. (\ref{output}) can also be
written as
\begin{equation}
\hat{a}_{p}^{(out)}(z)=T\hat{a}_{p}^{(in)}(0)  \label{output1}
\end{equation}
where
\begin{equation}
T=e^{-\frac{\chi ^{\prime \prime }\omega _{p}z}{2c}}e^{i\frac{\chi ^{\prime
}\omega _{p}z}{2c}}=e^{-\frac{\chi ^{\prime \prime }\omega _{p}t}{2}}e^{i%
\frac{\chi ^{\prime }\omega _{p}t}{2}}  \label{Transmission coeff}
\end{equation}
is the transmission coefficient for the vapor.

\section{SPDC in EIT}

SPDC has been studied for many years
\cite{Klyshko,Shih,SPDC,ShihAlley,Shih1,Todd,Stre,Rubin,Rubin1}.
In this section we shall revisit type-II SPDC, paying particular
attention to the passage of the beams through optical devices and
the detection of the photons. We will consider both the
single-photon counting detection and two-photon coincidence
counting detection.

\subsection{Single-Photon Detection of SPDC in EIT Medium}

\begin{figure}[tbp]
\includegraphics{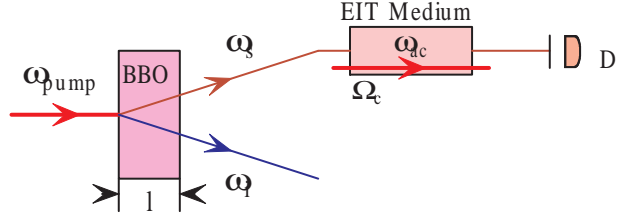}
\caption{Single-photon detection with EIT medium. $\omega _{ac}$
is the transition frequency of the medium between states
$|a\rangle $ and $|c\rangle $. $\Omega _{c}$ represents the
coupling field. $l$ is the length of the crystal.} \label{Fig.2}
\end{figure}

To fix notation we first discuss the single-photon properties of the
radiation. Consider the experiment illustrated in Fig. \ref{Fig.2}, a
monochromatic laser beam (pump beam with frequency $\omega _{pump}$)
incident on a noncentrosymmetric birefringent crystal (e.g. BBO) produces
pairs of photons. In Fig. \ref{Fig.2} the point detector D detects the
signal beam after EIT medium.

The average single-photon counting rate at detector D with efficiency $\eta $
is given by:
\begin{equation}
R_{s}=\eta \int\limits_{0}^{T}{dt_{s}}\langle \Psi |\hat{E}_{s}^{(-)}\hat{E}%
_{s}^{(+)}|\Psi\rangle  \label{SingleCR}
\end{equation}
The field $\hat{E}_{s}^{(+)}$ is the positive frequency part of the signal
field evaluated at the position $\vec{r}_{s}$ and the time $t_{s}$ and $\hat{%
E}_{s}^{(-)}$ is its Hermitian conjugate. $|\Psi\rangle$ is the
state of the system at the output surface of the crystal.
Generally $|\Psi\rangle$ is a superposition of the vacuum state
$|0\rangle$ and states with any number of pairs of photons.
Because of the small nonlinearity of the crystal, the expansion of
$|\Psi\rangle$ in the perturbation theory is limited to the first
two terms,
\begin{equation}
|\Psi\rangle=|0\rangle+\sum\limits_{\vec{k}_{s},\vec{k}_{i}}\Pi (\vec{k}_{s},\vec{k}%
_{i})\hat{a}_{s}^{+}(\vec{k}_{s})\hat{a}_{i}^{+}(\vec{k}_{i})|0\rangle
\label{Biphoton}
\end{equation}
where $\vec{k}_{s}$ and $\vec{k}_{i}$ are the wave vectors of the signal and
idler inside the crystal and $\hat{a}_{j}^{+}$ ($j=s,i$) is the creation
operator at the surface of the crystal. $\Pi (\vec{k}_{s},\vec{k}_{i})$ is
the spectral function of the two-photon state determined from phase
matching. The general form of the spectral function is
\begin{equation}
\Pi (\vec{k}_{s},\vec{k}_{i})=F(W_{s},W_{i})\Phi (\nu DL)h_{tr}(\vec{q}_{s},%
\vec{q}_{i})\delta (\omega _{s}+\omega _{i}-\omega _{p})  \label{SpectralF}
\end{equation}
where
\begin{equation}
D=\frac{1}{v_{s}}-\frac{1}{v_{i}},  \label{D}
\end{equation}
is the difference of inverse group velocities of the signal and
idler, and the Dirac $\delta $ function arises from the
steady-state or frequency phase-matching condition. The phase
matching condition in the
transverse direction is determined by the function $h_{tr}(\vec{q}_{s},\vec{q}%
_{i})$ where $\vec{q}_{s}$ and $\vec{q}_{i}$ are the transverse components
of wave vectors of the signal and idler, respectively. The longitudinal
phase matching condition gives $\Phi (\nu Dl)$ where $l$ is the length of
the crystal and $\nu $ $=\omega _{s}-\omega _{i}.$ Finally, all of slowly
varying variables are absorbed into $F(W_{s},W_{i})$ and $W_{s}$ and $W_{i}$
are the central frequencies of the signal and idler.

For simplicity, the following discussions are focused on the collinear case
of degenerate type-II SPDC, i.e., the propagation along $z$ axis and $%
W_{s}=W_{i}=\omega _{p}/2$. In this special case, the state of the system
takes the form of \cite{Shih}
\begin{equation}
|\Psi \rangle=\int {d\omega _{s}d\omega _{i}}\Phi (\nu Dl)\delta
(\omega _{s}+\omega _{i}-\omega
_{p})\hat{a}_{s}^{+}\hat{a}_{i}^{+}|0 \rangle \label{OneDim}
\end{equation}
with the phase matching function
\begin{equation}
\Phi (\nu Dl)=\frac{1-e^{i\nu DL}}{i\nu Dl}.  \label{Phi function}
\end{equation}
Note here we ignore all of constants and slowly varying variables in Eq. (%
\ref{OneDim}).

\begin{figure}[tbp]
\includegraphics{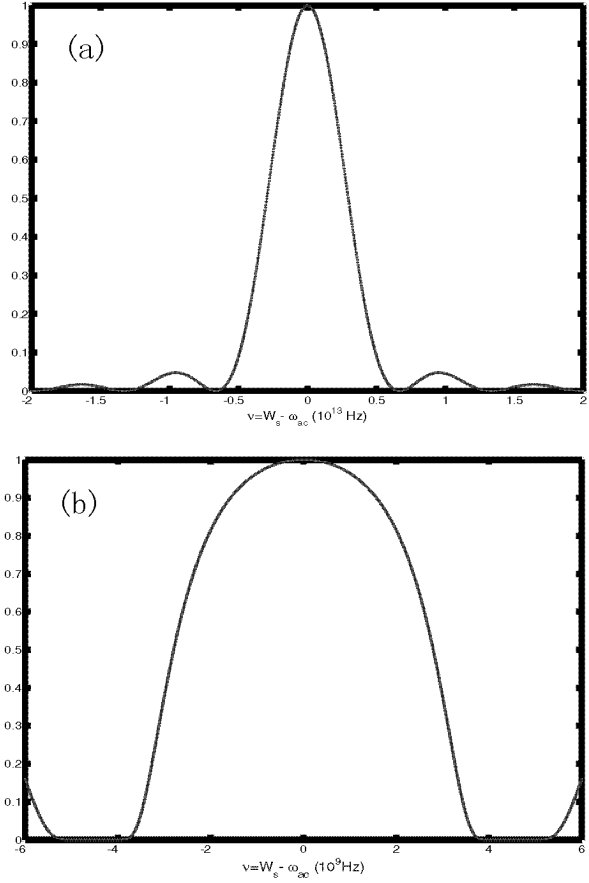}
\caption{(a) Spectrum of Single-photon counts in degenerate type-II SPDC
without an EIT medium. (b) Spectrum of single-photon counts in degenerate
type-II SPDC with an EIT medium. Here the central frequency $W_{s}$ of the
signal coinciding with the transition rate $\omega _{ac}$ is assumed, i.e., $%
W_{s}=\omega _{ac}$.}
\label{Fig.3}
\end{figure}

For the CW pump case considered here the single-photon counting rate is
constant and is given by
\begin{equation}
R_{s}=\eta \int {d\nu }|\Phi (\nu Dl)T|^{2}=\eta \int {d\nu }S(\nu )
\label{SingleCR1}
\end{equation}
where $T$ is the transmission coefficient (Eq. \ref{Transmission
coeff}),
\begin{equation}
S(\nu )=S_{0}(\nu )e^{\frac{\chi ^{\prime \prime }\omega _{p}L}{v_{g}}},
\label{S singles spectral f}
\end{equation}
and the unfiltered spectral function is
\begin{equation}
S_{0}(\nu )=\text{sinc}^{2}(\frac{\nu Dl}{2}).  \label{sinc}
\end{equation}
The comparison between these two cases is illustrated in Fig. \ref{Fig.3}.
Note here that there are two frequency differences: one is coming
from the non-perfect phase matching condition of SPDC, $\nu ,$ and
the other is from frequency detuning of the signal field in the
EIT medium, $\Delta =W_{s}+\nu -\omega _{ac}$. In Fig. \ref{Fig.3}
$S_{0}(\nu )$ and $S(\nu )$ are plotted. Note the different
scales, from which we see that the dominate effect is the EIT
absorption profile. The two sharp ''dips'' are the signature of
EIT phenomena, and correspond to two absorption peaks of the EIT
medium. The interval between those two dips is proportional to the
bandwidth of transparency window. If the central frequency of the
signal beam does not coincide with the atomic transition rate,
$W_{s}\neq \omega _{ac}$, the spectrum distribution becomes
asymmetric and the central peak will shift to
left or right, which is determined by the larger one between $W_{s}$ and $%
\omega _{ac}$. In order to prevent the bulk of the signal photons from
obscuring the EIT signal it would be necessary to filter the beam using a
narrow filter at the input of the cell.

\subsection{Two-Photon Interference of SPDC in EIT Medium}

\begin{figure}[tbp]
\includegraphics{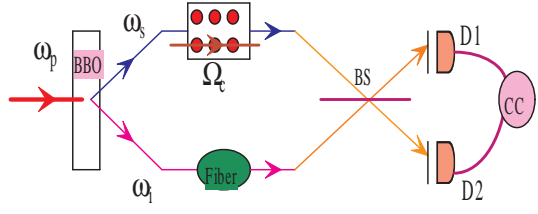}
\caption{Two-photon coincidence counting detection with EIT medium. A
birefringent crystal (BBO) converts the pump beam into a pair of
down-converted beams. BS is a 50-50 beam splitter, D1 and D2 are
photodetectors, and CC is the coincidence counter.}
\label{Fig.4}
\end{figure}

In order to measure the two-photon correlation let us consider the
simplified experiment shown in Fig. \ref{Fig.4} \cite{ShihAlley}. The
entangled signal and idler photon pair emitted in the SPDC process is mixed
by a 50-50 beam splitter (BS) and then recorded by photodetectors D1 and D2
for coincidence. In the idler channel, a time-delay apparatus ($\delta \tau $%
) is put to balance the signal and idler path-lengths.

The two-photon coincidence counting rate is defined as \cite{Klyshko,Shih}:
\begin{eqnarray}
R_{cc} &=&\eta ^{2}\frac{1}{T}\int\limits_{0}^{T}{dt_{1}}\int\limits_{0}^{T}{%
dt_{2}}\langle \Psi |\hat{E}_{1}^{(-)}\hat{E}_{2}^{(-)}\hat{E}_{2}^{(+)}\hat{%
E}_{1}^{(+)}|\Psi \rangle   \nonumber \\
&=&\eta ^{2}\frac{1}{T}\int\limits_{0}^{T}{dt_{1}}\int\limits_{0}^{T}{dt_{2}}%
|\Psi (t_{1},t_{2})|^{2},  \label{Coincidence}
\end{eqnarray}
where
\begin{equation}
\Psi (t_{1},t_{2})=\langle 0|\hat{E}_{1}^{(+)}\hat{E}_{2}^{(+)}|\Psi \rangle
\label{BiphotonAmp}
\end{equation}
is the effective two-photon amplitude. Taking into account the EIT cell, the
two-photon amplitude is
\begin{equation}
\Psi (t_{1},t_{2})=\int {d\nu }\Phi (\nu Dl)Te^{-i(W_{s}+\nu
)t_{1}}e^{-i(W_{i}-\nu )t_{2}}  \label{TwoPhotonWF}
\end{equation}
As a comparison, the two-photon amplitude without an EIT medium is given by:
\begin{equation}
\Psi _{0}(t_{1},t_{2})=\int {d\nu }\Phi (\nu Dl)e^{-i(W_{s}+\nu
)t_{1}}e^{-i(W_{i}-\nu )t_{2}}.  \label{TwoPhotonWFNoEIT}
\end{equation}
The coincidence counting rate for type-II SPDC now can be evaluated
\begin{eqnarray}
R_{cc} &=&\eta ^{2}\int {d\nu }\{|\Phi (\nu Dl)T(\nu )|^{2}-Re[\Phi ^{*}(\nu
Dl)  \nonumber \\
&&\times \Phi [(W_{i}-W_{s}-\nu )Dl]T^{*}(\nu )T(W_{i}-W_{s}-\nu )  \nonumber
\\
&&\times e^{-i(W_{s}-W_{i}+2\nu )\delta \tau }]\}.  \label{Rcc}
\end{eqnarray}

In the following discussions, we concentrate on the degenerate type-II SPDC,
then the counting rate becomes
\begin{eqnarray}
R_{cc} &=&Re\eta ^{2}\int {d\nu }\text{ sinc}^{2}(\frac{\nu
Dl}{2}) \times [|T(\nu )|^{2}-T^{*}(\nu )T(-\nu )e^{i\nu
(Dl-2\delta \tau )}]
\nonumber \\
&=&\eta ^{2}\int {d\nu }\text{ sinc}^{2}(\frac{\nu
Dl}{2})e^{-\frac{\chi ^{\prime \prime }\omega _{s}L}{c}} \times
\{1-\cos [\nu (Dl-2\delta \tau) -2 \phi_{d}]\}. \label{RccII}
\end{eqnarray}
where  $\phi _{d}(\nu)=\chi ^{\prime}W_{s}L/2c$, with $L$ the length of the EIT cell, is the 
phase delay of a pulse due to the EIT. We have
assumed that the dispersion in the fiber can be ignored. For conventional
two-photon interference experiments, removing the EIT medium in the signal
channel, the coincidence counts of degenerate type-II SPDC read
\begin{equation}
R_{cc}^{(0)}=\eta ^{2}\int {d\nu }|\Phi (\frac{\nu
Dl}{2})|^{2}[1-\cos \nu (Dl-2\delta \tau )]  \label{Rcc0}
\end{equation}
For simplicity, we assume perfect detection of photodetectors and 
ignored other losses,
\textit{i.e.}, taking $\eta=1$ in the following discussions.

\begin{figure}[tbp]
\includegraphics{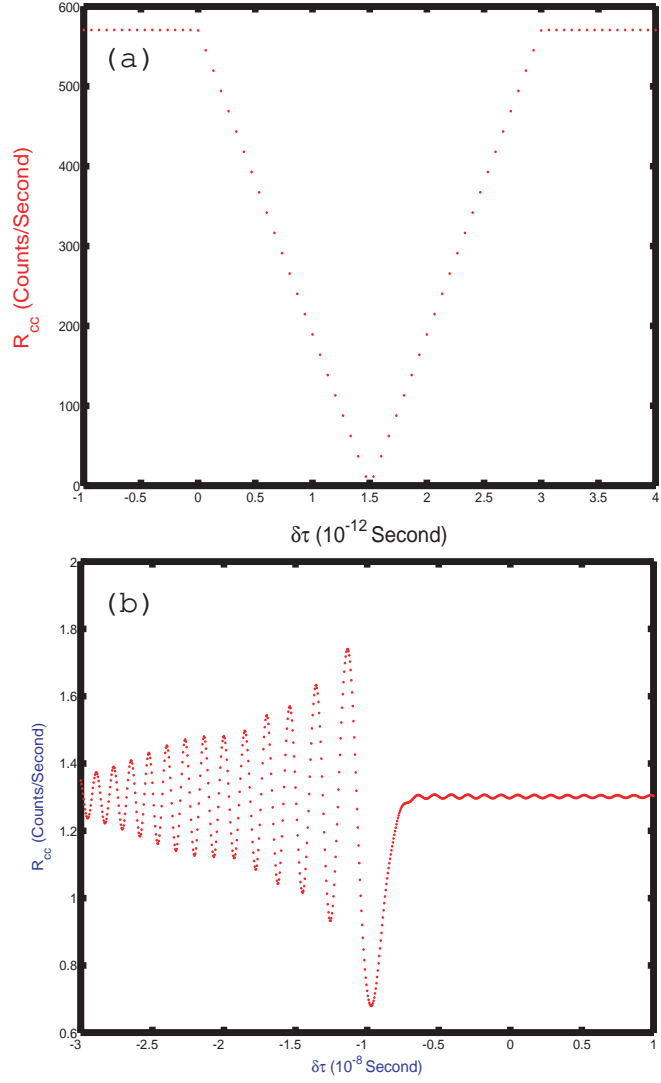}
\caption{Two-photon coincidence counting rates of degenerate type-II SPDC.
(a) conventional measurement. (b) with EIT medium $(Rb^{87})$. The
parameters used here are: $Dl=3\times {10}^{-12}Sec$, $L=0.1m$, $\Omega %
_{c}=2\protect\sqrt{5}\times {10}^{9}Hz$ (5.5 Watts), and $N=2\times {10}%
^{18}/m^{3}$ (SI units).}
\label{Fig.5}
\end{figure}

The comparison of coincidence counting rates between with the EIT medium and
without it is shown in Fig. \ref{Fig.5}, again note the different frequency
scales.
The quantum coherence from SPDC process still exists as is evident
from Fig. \ref{Fig.5} (b) where we see that in addition to a dip,
there is an oscillation. The dip profile is much narrower than the
notch shown in Fig. \ref{Fig.5} (a) with no EIT cell. Recall that
the center of the notch for the conventional coincidence counts is
determined by $Dl/2$, and  its width
is determined by the quantity $Dl$ (Eq. (\ref{Rcc0})). In Fig. \ref{Fig.5}%
(b), the dip center is displaced from $Dl/2$ by the time delay $\tau _{d}$
caused by EIT effect.  The oscillations can be understood as follows.  For
the calculation of the figure we chose the special case that the center
frequency of the signal beam coincides with the atomic transition rate,
i.e., $W_{s}=\omega _{ca}$. The transmission peak has a width of the   $%
\Delta \omega _{tr}$ and is centered on $\nu =0.$ If $\Delta \omega _{tr}%
\times (Dl-2\phi' _{d}-2\delta \tau )>1$ oscillations can occur while for  $%
\Delta \omega _{tr}\times (Dl-2\phi'_{d}-2\delta \tau )<1$, where 
$\phi_{d}'=d\phi_{d}/d\nu$ they
will disappear. For the parameters used here, the width of
transparency window is $\Delta \omega _{tr}=5.527\times 10^{9}Hz$,
the group velocity of single photons in the EIT medium is
$v_{g}=1.064\times 10^{7}m/s$, and the
corresponding time delay is $\tau _{d}=9.069\times 10^{-9}Sec$, so if $%
\delta \tau $ is negative we can get oscillations in the counting
rate. In addition, because the transparency bandwidth of EIT is
much narrower than the bandwidth of SPDC, the coincidence counting
rate with an EIT medium (Fig. \ref{Fig.5} (b)) is much lower than
the conventional case. For the parameters given above, ignoring
losses, the coincidence counts are around one photon per second,
which is barely doable in the lab.

\begin{figure}[tbp]
\includegraphics{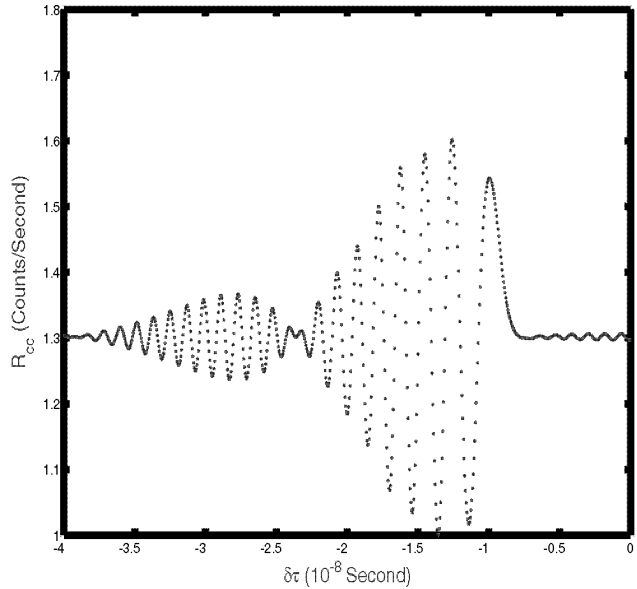}
\caption{Two-photon coincidence counting rates of degenerate type-II SPDC
with EIT medium $(Rb^{87})$ and small detuning. The parameters used here
are: $W_{s}-\omega _{ca}=10^{8}Hz$, $Dl=3\times {10}^{-12}Sec$, $L=0.1m$, $%
\Omega _{c}=2\protect\sqrt{5}\times {10}^{9}Hz$, and $N=2\times {10}%
^{18}/m^{3}$ (SI units).}
\label{Fig.6}
\end{figure}

In Fig. \ref{Fig.5}(b) the dip center coincides with the time
delay $\tau _{d}$ because of EIT dominant effect. If we do not use
the resonant condition for the probe $W_{s}=\omega _{ca}$, the
visibility of coincidence counts is lowered and meanwhile the
oscillations are modulated by the frequency mismatching (see Fig.
\ref{Fig.6}). The visibility can be enhanced or decreased
depending on the sign and magnitude of that frequency mismatching.

\section{Conclusion}

In this paper we present the methods to consider light storage at a
single-photon level. The relation between the input and output field
operator with an EIT medium is derived. Based on this relationship, the
non-classical light (SPDC) storage is studied. We focused on the discussions
about single-photon counts and two-photon coincidence counts of degenerate
type-II SPDC with an EIT medium in slow light case. Though the experiment
may be hard to implement, it is doable depending on the choice of
experimental parameters. The results we have obtained are corresponding to
the assumption that the coupling field is not a function of time.

\begin{acknowledgments}

We would like to thank Michael Steiner for pointing out the
importance of filtering the input beam.  We also thank Prof. Yanhua Shih,
the members of Quantum
Optics Group at UMBC, and our collaborators at NRL for useful
discussions. This work was partly supported by NRL through Grant
No. code 5312.

\end{acknowledgments}

\end{document}